\documentstyle[12pt]{article}

\newskip\humongous \humongous=0pt plus 1000pt minus 1000pt

\newif\ifdtup

\def\VEV#1{\left\langle #1\right\rangle}

\def\ltap{\;\raisebox{-.4ex}{\rlap{$\sim$}} \raisebox{.4ex}{$<$}\;}
\def\gtap{\;\raisebox{-.4ex}{\rlap{$\sim$}} \raisebox{.4ex}{$>$}\;}

\def\beq{\begin{equation}}
\def\eeq{\end{equation}}
\def\eq{\beq\eeq}
\def\beqn{\begin{eqnarray}}
\def\eeqn{\end{eqnarray}}
\relax

\def\dotx{\dotx{\dot\overline{x}}}

\relax

\catcode`\@=11

\@addtoreset{equation}{section}
\def\theequation{\thesection\arabic{equation}}

\def\@normalsize{\@setsize\normalsize{15pt}\xiipt\@xiipt
\abovedisplayskip 14pt plus3pt minus3pt%
\belowdisplayskip \abovedisplayskip
\abovedisplayshortskip \z@ plus3pt%
\belowdisplayshortskip 7pt plus3.5pt minus0pt}

\def\small{\@setsize\small{13.6pt}\xipt\@xipt
\abovedisplayskip 13pt plus3pt minus3pt%
\belowdisplayskip \abovedisplayskip
\abovedisplayshortskip \z@ plus3pt%
\belowdisplayshortskip 7pt plus3.5pt minus0pt
\def\@listi{\parsep 4.5pt plus 2pt minus 1pt
     \itemsep \parsep
     \topsep 9pt plus 3pt minus 3pt}}

\@twosidetrue

\relax

\catcode`@=12

\catcode`\@=11

\def\section{\@startsection{section}{1}{\z@}{3.5ex plus 1ex minus
   .2ex}{2.3ex plus .2ex}{\large\bf}}

\def\thesection{\arabic{section}.}

\def\appendix{\setcounter{section}{0}
 \def\thesection{APPENDIX \Alph{section}:}
 \def\theequation{\Alph{section}.\arabic{equation}}}

\def\ps@headings{\def\@oddfoot{}\def\@evenfoot{}
\def\@oddhead{\hbox{}\hfill
 \makebox[.5\textwidth]{\raggedright\ignorespaces --\thepage{}--
 \hfill {}}}  
\def\@evenhead{\@oddhead}
\def\subsectionmark##1{\markboth{##1}{}}
}

\ps@headings

\catcode`\@=12

\def\figcap{\section*{Figure Captions\markboth
 {FIGURECAPTIONS}{FIGURECAPTIONS}}\list
 {Fig. \arabic{enumi}:\hfill}{\settowidth\labelwidth{Fig. 999:}
 \leftmargin\labelwidth
 \advance\leftmargin\labelsep\usecounter{enumi}}}
 \relax
\def\tablecap{\section*{Table Captions\markboth
 {TABLECAPTIONS}{TABLECAPTIONS}}\list
 {Table \arabic{enumi}:\hfill}{\settowidth\labelwidth{Table 999:}
 \leftmargin\labelwidth
 \advance\leftmargin\labelsep\usecounter{enumi}}}
 \relax
\def\reflist{\section*{References\markboth
 {REFLIST}{REFLIST}}\list
 {[\arabic{enumi}]\hfill}{\settowidth\labelwidth{[999]}
 \leftmargin\labelwidth
 \advance\leftmargin\labelsep\usecounter{enumi}}}
 \relax

\catcode`\@=11

\def\ps@headings{\def\@oddfoot{}\def\@evenfoot{}
\def\@oddhead{\hbox{}\hfill
 \makebox[.5\textwidth]{\raggedright\ignorespaces --\thepage{}--
 \hfill {}}}    
\def\@evenhead{\@oddhead}
\def\subsectionmark##1{\markboth{##1}{}}
}

\ps@headings

\relax

\relax
\def\pl#1#2#3{{\it Phys. Lett. }{\bf #1}(19#2)#3}
\def\zp#1#2#3{{\it Z. Phys. }{\bf #1}(19#2)#3}

\def\np#1#2#3{{\it Nucl. Phys. }{\bf #1}(19#2)#3}

\relax

\ifx\undefined\psfig\else\endinput\fi

\edef\psfigRestoreAt{\catcode`@=\number\catcode`@\relax}
\catcode`\@=11\relax
\newwrite\@unused
\def\ps@typeout#1{{\let\protect\string\immediate\write\@unused{#1}}}
\ps@typeout{psfig/tex 1.8}

\def\figurepath{./}

\def\@nnil{\@nil}
\def\@empty{}
\def\@psdonoop#1\@@#2#3{}
\def\@psdo#1:=#2\do#3{\edef\@psdotmp{#2}\ifx\@psdotmp\@empty \else
    \expandafter\@psdoloop#2,\@nil,\@nil\@@#1{#3}\fi}
\def\@psdoloop#1,#2,#3\@@#4#5{\def#4{#1}\ifx #4\@nnil \else
       #5\def#4{#2}\ifx #4\@nnil \else#5\@ipsdoloop #3\@@#4{#5}\fi\fi}
\def\@ipsdoloop#1,#2\@@#3#4{\def#3{#1}\ifx #3\@nnil
       \let\@nextwhile=\@psdonoop \else
      #4\relax\let\@nextwhile=\@ipsdoloop\fi\@nextwhile#2\@@#3{#4}}
\def\@tpsdo#1:=#2\do#3{\xdef\@psdotmp{#2}\ifx\@psdotmp\@empty \else
    \@tpsdoloop#2\@nil\@nil\@@#1{#3}\fi}
\def\@tpsdoloop#1#2\@@#3#4{\def#3{#1}\ifx #3\@nnil
       \let\@nextwhile=\@psdonoop \else
      #4\relax\let\@nextwhile=\@tpsdoloop\fi\@nextwhile#2\@@#3{#4}}
\ifx\undefined\fbox
\newdimen\fboxrule
\newdimen\fboxsep
\newdimen\ps@tempdima
\newbox\ps@tempboxa
\long\def\fbox#1{\leavevmode\setbox\ps@tempboxa\hbox{#1}\ps@tempdima\fboxrule
    \advance\ps@tempdima \fboxsep \advance\ps@tempdima \dp\ps@tempboxa
   \hbox{\lower \ps@tempdima\hbox
  {\vbox{\hrule height \fboxrule
          \hbox{\vrule width \fboxrule \hskip\fboxsep
          \vbox{\vskip\fboxsep \box\ps@tempboxa\vskip\fboxsep}\hskip
                 \fboxsep\vrule width \fboxrule}
                 \hrule height \fboxrule}}}}
\fi
\newread\ps@stream
\newif\ifnot@eof       
\newif\if@noisy        
\newif\if@atend        
\newif\if@psfile       
{\catcode`\%=12\global\gdef\epsf@start{
\def\epsf@PS{PS}
\def\epsf@getbb#1{%
\openin\ps@stream=#1
\ifeof\ps@stream\ps@typeout{Error, File #1 not found}\else
   {\not@eoftrue \chardef\other=12
    \def\do##1{\catcode`##1=\other}\dospecials \catcode`\ =10
    \loop
       \if@psfile
	  \read\ps@stream to \epsf@fileline
       \else{
	  \obeyspaces
          \read\ps@stream to \epsf@tmp\global\let\epsf@fileline\epsf@tmp}
       \fi
       \ifeof\ps@stream\not@eoffalse\else
       \if@psfile\else
       \expandafter\epsf@test\epsf@fileline:. \\%
       \fi
          \expandafter\epsf@aux\epsf@fileline:. \\%
       \fi
   \ifnot@eof\repeat
   }\closein\ps@stream\fi}%
\long\def\epsf@test#1#2#3:#4\\{\def\epsf@testit{#1#2}
			\ifx\epsf@testit\epsf@start\else
\ps@typeout{Warning! File does not start with `\epsf@start'.  It may not be a
PostScript file.}
			\fi
			\@psfiletrue} 
{\catcode`\%=12\global\let\epsf@percent=
\long\def\epsf@aux#1#2:#3\\{\ifx#1\epsf@percent
   \def\epsf@testit{#2}\ifx\epsf@testit\epsf@bblit
	\@atendfalse
        \epsf@atend #3 . \\%
	\if@atend
	   \if@verbose{
		\ps@typeout{psfig: found `(atend)'; continuing search}
	   }\fi
        \else
        \epsf@grab #3 . . . \\%
        \not@eoffalse
        \global\no@bbfalse
        \fi
   \fi\fi}%
\def\epsf@grab #1 #2 #3 #4 #5\\{%
   \global\def\epsf@llx{#1}\ifx\epsf@llx\empty
      \epsf@grab #2 #3 #4 #5 .\\\else
   \global\def\epsf@lly{#2}%
   \global\def\epsf@urx{#3}\global\def\epsf@ury{#4}\fi}%
\def\epsf@atendlit{(atend)}
\def\epsf@atend #1 #2 #3\\{%
   \def\epsf@tmp{#1}\ifx\epsf@tmp\empty
      \epsf@atend #2 #3 .\\\else
   \ifx\epsf@tmp\epsf@atendlit\@atendtrue\fi\fi}

\chardef\letter = 11
\chardef\other = 12

\newif \ifdebug 
\newif\ifc@mpute 
\c@mputetrue 

\let\then = \relax
\def\r@dian{pt }
\let\r@dians = \r@dian
\let\dimensionless@nit = \r@dian
\let\dimensionless@nits = \dimensionless@nit
\def\internal@nit{sp }
\let\internal@nits = \internal@nit
\newif\ifstillc@nverging
\def \Mess@ge #1{\ifdebug \then \message {#1} \fi}

{ 
	\catcode `\@ = \letter
	\gdef \nodimen {\expandafter \n@dimen \the \dimen}
	\gdef \term #1 #2 #3%
	       {\edef \t@ {\the #1}
		\edef \t@@ {\expandafter \n@dimen \the #2\r@dian}%
		\t@rm {\t@} {\t@@} {#3}%
	       }
	\gdef \t@rm #1 #2 #3%
	       {{%
		\count 0 = 0
		\dimen 0 = 1 \dimensionless@nit
		\dimen 2 = #2\relax
		\Mess@ge {Calculating term #1 of \nodimen 2}%
		\loop
		\ifnum	\count 0 < #1
		\then	\advance \count 0 by 1
			\Mess@ge {Iteration \the \count 0 \space}%
			\Multiply \dimen 0 by {\dimen 2}%
			\Mess@ge {After multiplication, term = \nodimen 0}%
			\Divide \dimen 0 by {\count 0}%
			\Mess@ge {After division, term = \nodimen 0}%
		\repeat
		\Mess@ge {Final value for term #1 of
				\nodimen 2 \space is \nodimen 0}%
		\xdef \Term {#3 = \nodimen 0 \r@dians}%
		\aftergroup \Term
	       }}
	\catcode `\p = \other
	\catcode `\t = \other
	\gdef \n@dimen #1pt{#1} 
}

\def \Divide #1by #2{\divide #1 by #2} 

\def \Multiply #1by #2
       {{
	\count 0 = #1\relax
	\count 2 = #2\relax
	\count 4 = 65536
	\Mess@ge {Before scaling, count 0 = \the \count 0 \space and
			count 2 = \the \count 2}%
	\ifnum	\count 0 > 32767 
	\then	\divide \count 0 by 4
		\divide \count 4 by 4
	\else	\ifnum	\count 0 < -32767
		\then	\divide \count 0 by 4
			\divide \count 4 by 4
		\else
		\fi
	\fi
	\ifnum	\count 2 > 32767 
	\then	\divide \count 2 by 4
		\divide \count 4 by 4
	\else	\ifnum	\count 2 < -32767
		\then	\divide \count 2 by 4
			\divide \count 4 by 4
		\else
		\fi
	\fi
	\multiply \count 0 by \count 2
	\divide \count 0 by \count 4
	\xdef \product {#1 = \the \count 0 \internal@nits}%
	\aftergroup \product
       }}

\def\r@duce{\ifdim\dimen0 > 90\r@dian \then   
		\multiply\dimen0 by -1
		\advance\dimen0 by 180\r@dian
		\r@duce
	    \else \ifdim\dimen0 < -90\r@dian \then  
		\advance\dimen0 by 360\r@dian
		\r@duce
		\fi
	    \fi}

\def\Sine#1%
       {{%
	\dimen 0 = #1 \r@dian
	\r@duce
	\ifdim\dimen0 = -90\r@dian \then
	   \dimen4 = -1\r@dian
	   \c@mputefalse
	\fi
	\ifdim\dimen0 = 90\r@dian \then
	   \dimen4 = 1\r@dian
	   \c@mputefalse
	\fi
	\ifdim\dimen0 = 0\r@dian \then
	   \dimen4 = 0\r@dian
	   \c@mputefalse
	\fi
	\ifc@mpute \then
		\divide\dimen0 by 180
		\dimen0=3.141592654\dimen0
		\dimen 2 = 3.1415926535897963\r@dian 
		\divide\dimen 2 by 2 
		\Mess@ge {Sin: calculating Sin of \nodimen 0}%
		\count 0 = 1 
		\dimen 2 = 1 \r@dian 
		\dimen 4 = 0 \r@dian 
		\loop
			\ifnum	\dimen 2 = 0 
			\then	\stillc@nvergingfalse
			\else	\stillc@nvergingtrue
			\fi
			\ifstillc@nverging 
			\then	\term {\count 0} {\dimen 0} {\dimen 2}%
				\advance \count 0 by 2
				\count 2 = \count 0
				\divide \count 2 by 2
				\ifodd	\count 2 
				\then	\advance \dimen 4 by \dimen 2
				\else	\advance \dimen 4 by -\dimen 2
				\fi
		\repeat
	\fi
			\xdef \sine {\nodimen 4}%
       }}

\def\Cosine#1{\ifx\sine\UnDefined\edef\Savesine{\relax}\else
		             \edef\Savesine{\sine}\fi
	{\dimen0=#1\r@dian\advance\dimen0 by 90\r@dian
	 \Sine{\nodimen 0}
	 \xdef\cosine{\sine}
	 \xdef\sine{\Savesine}}}

\def\psdraft{
	\def\@psdraft{0}
}
\def\psfull{
	\def\@psdraft{100}
}

\psfull

\newif\if@scalefirst
\def\psscalefirst{\@scalefirsttrue}
\def\psrotatefirst{\@scalefirstfalse}
\psrotatefirst

\newif\if@draftbox
\def\psnodraftbox{
	\@draftboxfalse
}
\def\psdraftbox{
	\@draftboxtrue
}
\@draftboxtrue

\newif\if@prologfile
\newif\if@postlogfile
\def\pssilent{
	\@noisyfalse
}
\def\psnoisy{
	\@noisytrue
}
\psnoisy
\newif\if@bbllx
\newif\if@bblly
\newif\if@bburx
\newif\if@bbury
\newif\if@height
\newif\if@width
\newif\if@rheight
\newif\if@rwidth
\newif\if@angle
\newif\if@clip
\newif\if@verbose
\def\@p@@sclip#1{\@cliptrue}

\newif\if@decmpr

\def\@p@@sfigure#1{\def\@p@sfile{null}\def\@p@sbbfile{null}
	        \openin1=#1.bb
		\ifeof1\closein1
	        	\openin1=\figurepath#1.bb
			\ifeof1\closein1
			        \openin1=#1
				\ifeof1\closein1%
				       \openin1=\figurepath#1
					\ifeof1
					   \typeout{Error, File #1 not found}
						\if@bbllx\if@bblly
				   		\if@bburx\if@bbury
			      				\def\@p@sfile{#1}%
			      				\def\@p@sbbfile{#1}%
							\@decmprfalse
				  	   	\fi\fi\fi\fi
					\else\closein1
				    		\def\@p@sfile{\figurepath#1}%
				    		\def\@p@sbbfile{\figurepath#1}%
						\@decmprfalse
	                       		\fi%
			 	\else\closein1%
					\def\@p@sfile{#1}
					\def\@p@sbbfile{#1}
					\@decmprfalse
			 	\fi
			\else
				\def\@p@sfile{\figurepath#1}
				\def\@p@sbbfile{\figurepath#1.bb}
				\@decmprtrue
			\fi
		\else
			\def\@p@sfile{#1}
			\def\@p@sbbfile{#1.bb}
			\@decmprtrue
		\fi}

\def\@p@@sfile#1{\@p@@sfigure{#1}}

\def\@p@@sbbllx#1{
		\@bbllxtrue
		\dimen100=#1
		\edef\@p@sbbllx{\number\dimen100}
}
\def\@p@@sbblly#1{
		\@bbllytrue
		\dimen100=#1
		\edef\@p@sbblly{\number\dimen100}
}
\def\@p@@sbburx#1{
		\@bburxtrue
		\dimen100=#1
		\edef\@p@sbburx{\number\dimen100}
}
\def\@p@@sbbury#1{
		\@bburytrue
		\dimen100=#1
		\edef\@p@sbbury{\number\dimen100}
}
\def\@p@@sheight#1{
		\@heighttrue
		\dimen100=#1
   		\edef\@p@sheight{\number\dimen100}
}
\def\@p@@swidth#1{
		\@widthtrue
		\dimen100=#1
		\edef\@p@swidth{\number\dimen100}
}
\def\@p@@srheight#1{
		\@rheighttrue
		\dimen100=#1
		\edef\@p@srheight{\number\dimen100}
}
\def\@p@@srwidth#1{
		\@rwidthtrue
		\dimen100=#1
		\edef\@p@srwidth{\number\dimen100}
}
\def\@p@@sangle#1{
		\@angletrue
		\edef\@p@sangle{#1} 
}
\def\@p@@ssilent#1{
		\@verbosefalse
}
\def\@p@@sprolog#1{\@prologfiletrue\def\@prologfileval{#1}}
\def\@p@@spostlog#1{\@postlogfiletrue\def\@postlogfileval{#1}}
\def\@cs@name#1{\csname #1\endcsname}
\def\@setparms#1=#2,{\@cs@name{@p@@s#1}{#2}}
\def\ps@init@parms{
		\@bbllxfalse \@bbllyfalse
		\@bburxfalse \@bburyfalse
		\@heightfalse \@widthfalse
		\@rheightfalse \@rwidthfalse
		\def\@p@sbbllx{}\def\@p@sbblly{}
		\def\@p@sbburx{}\def\@p@sbbury{}
		\def\@p@sheight{}\def\@p@swidth{}
		\def\@p@srheight{}\def\@p@srwidth{}
		\def\@p@sangle{0}
		\def\@p@sfile{} \def\@p@sbbfile{}
		\def\@p@scost{10}
		\def\@sc{}
		\@prologfilefalse
		\@postlogfilefalse
		\@clipfalse
		\if@noisy
			\@verbosetrue
		\else
			\@verbosefalse
		\fi
}
\def\parse@ps@parms#1{
	 	\@psdo\@psfiga:=#1\do
		   {\expandafter\@setparms\@psfiga,}}
\newif\ifno@bb
\def\bb@missing{
	\if@verbose{
		\typeout{psfig: searching \@p@sbbfile \space  for bounding box}
	}\fi
	\no@bbtrue
	\epsf@getbb{\@p@sbbfile}
        \ifno@bb \else \bb@cull\epsf@llx\epsf@lly\epsf@urx\epsf@ury\fi
}
\def\bb@cull#1#2#3#4{
	\dimen100=#1 bp\edef\@p@sbbllx{\number\dimen100}
	\dimen100=#2 bp\edef\@p@sbblly{\number\dimen100}
	\dimen100=#3 bp\edef\@p@sbburx{\number\dimen100}
	\dimen100=#4 bp\edef\@p@sbbury{\number\dimen100}
	\no@bbfalse
}
\newdimen\p@intvaluex
\newdimen\p@intvaluey
\def\rotate@#1#2{{\dimen0=#1 sp\dimen1=#2 sp
		  \global\p@intvaluex=\cosine\dimen0
		  \dimen3=\sine\dimen1
		  \global\advance\p@intvaluex by -\dimen3
		  \global\p@intvaluey=\sine\dimen0
		  \dimen3=\cosine\dimen1
		  \global\advance\p@intvaluey by \dimen3
		  }}
\def\compute@bb{
		\no@bbfalse
		\if@bbllx \else \no@bbtrue \fi
		\if@bblly \else \no@bbtrue \fi
		\if@bburx \else \no@bbtrue \fi
		\if@bbury \else \no@bbtrue \fi
		\ifno@bb \bb@missing \fi
		\ifno@bb \ps@typeout{FATAL ERROR: no bb supplied or found}
			\no-bb-error
		\fi
		\count203=\@p@sbburx
		\count204=\@p@sbbury
		\advance\count203 by -\@p@sbbllx
		\advance\count204 by -\@p@sbblly
		\edef\ps@bbw{\number\count203}
		\edef\ps@bbh{\number\count204}
		\if@angle
			\Sine{\@p@sangle}\Cosine{\@p@sangle}
	        	{\dimen100=\maxdimen\xdef\r@p@sbbllx{\number\dimen100}
					    \xdef\r@p@sbblly{\number\dimen100}
			                    \xdef\r@p@sbburx{-\number\dimen100}
					    \xdef\r@p@sbbury{-\number\dimen100}}
                        \def\minmaxtest{
			   \ifnum\number\p@intvaluex<\r@p@sbbllx
			      \xdef\r@p@sbbllx{\number\p@intvaluex}\fi
			   \ifnum\number\p@intvaluex>\r@p@sbburx
			      \xdef\r@p@sbburx{\number\p@intvaluex}\fi
			   \ifnum\number\p@intvaluey<\r@p@sbblly
			      \xdef\r@p@sbblly{\number\p@intvaluey}\fi
			   \ifnum\number\p@intvaluey>\r@p@sbbury
			      \xdef\r@p@sbbury{\number\p@intvaluey}\fi
			   }
			\rotate@{\@p@sbbllx}{\@p@sbblly}
			\minmaxtest
			\rotate@{\@p@sbbllx}{\@p@sbbury}
			\minmaxtest
			\rotate@{\@p@sbburx}{\@p@sbblly}
			\minmaxtest
			\rotate@{\@p@sbburx}{\@p@sbbury}
			\minmaxtest
			\edef\@p@sbbllx{\r@p@sbbllx}\edef\@p@sbblly{\r@p@sbblly}
			\edef\@p@sbburx{\r@p@sbburx}\edef\@p@sbbury{\r@p@sbbury}
		\fi
		\count203=\@p@sbburx
		\count204=\@p@sbbury
		\advance\count203 by -\@p@sbbllx
		\advance\count204 by -\@p@sbblly
		\edef\@bbw{\number\count203}
		\edef\@bbh{\number\count204}
}
\def\in@hundreds#1#2#3{\count240=#2 \count241=#3
		     \count100=\count240	
		     \divide\count100 by \count241
		     \count101=\count100
		     \multiply\count101 by \count241
		     \advance\count240 by -\count101
		     \multiply\count240 by 10
		     \count101=\count240	
		     \divide\count101 by \count241
		     \count102=\count101
		     \multiply\count102 by \count241
		     \advance\count240 by -\count102
		     \multiply\count240 by 10
		     \count102=\count240	
		     \divide\count102 by \count241
		     \count200=#1\count205=0
		     \count201=\count200
			\multiply\count201 by \count100
		 	\advance\count205 by \count201
		     \count201=\count200
			\divide\count201 by 10
			\multiply\count201 by \count101
			\advance\count205 by \count201
		     \count201=\count200
			\divide\count201 by 100
			\multiply\count201 by \count102
			\advance\count205 by \count201
		     \edef\@result{\number\count205}
}
\def\compute@wfromh{
		\in@hundreds{\@p@sheight}{\@bbw}{\@bbh}
		\edef\@p@swidth{\@result}
}
\def\compute@hfromw{
	        \in@hundreds{\@p@swidth}{\@bbh}{\@bbw}
		\edef\@p@sheight{\@result}
}
\def\compute@handw{
		\if@height
			\if@width
			\else
				\compute@wfromh
			\fi
		\else
			\if@width
				\compute@hfromw
			\else
				\edef\@p@sheight{\@bbh}
				\edef\@p@swidth{\@bbw}
			\fi
		\fi
}
\def\compute@resv{
		\if@rheight \else \edef\@p@srheight{\@p@sheight} \fi
		\if@rwidth \else \edef\@p@srwidth{\@p@swidth} \fi
}
\def\compute@sizes{
	\compute@bb
	\if@scalefirst\if@angle
	\if@width
	   \in@hundreds{\@p@swidth}{\@bbw}{\ps@bbw}
	   \edef\@p@swidth{\@result}
	\fi
	\if@height
	   \in@hundreds{\@p@sheight}{\@bbh}{\ps@bbh}
	   \edef\@p@sheight{\@result}
	\fi
	\fi\fi
	\compute@handw
	\compute@resv}

\def\psfig#1{\vbox {
	\ps@init@parms
	\parse@ps@parms{#1}
	\compute@sizes
	\ifnum\@p@scost<\@psdraft{
		\special{ps::[begin] 	\@p@swidth \space \@p@sheight \space
				\@p@sbbllx \space \@p@sbblly \space
				\@p@sbburx \space \@p@sbbury \space
				startTexFig \space }
		\if@angle
			\special {ps:: \@p@sangle \space rotate \space}
		\fi
		\if@clip{
			\if@verbose{
				\ps@typeout{(clip)}
			}\fi
			\special{ps:: doclip \space }
		}\fi
		\if@prologfile
		    \special{ps: plotfile \@prologfileval \space } \fi
		\if@decmpr{
			\if@verbose{
				\typeout{psfig: including \@p@sfile.Z \space }
			}\fi
			\special{ps: plotfile "`zcat \@p@sfile.Z" \space }
		}\else{
			\if@verbose{
				\typeout{psfig: including \@p@sfile \space }
			}\fi
			\special{ps: plotfile \@p@sfile \space }
		}\fi
		\if@postlogfile
		    \special{ps: plotfile \@postlogfileval \space } \fi
		\special{ps::[end] endTexFig \space }
		\vbox to \@p@srheight true sp{
			\hbox to \@p@srwidth true sp{
				\hss
			}
		\vss
		}
	}\else{
		\if@draftbox{
			\hbox{\frame{\vbox to \@p@srheight true sp{
			\vss
			\hbox to \@p@srwidth true sp{ \hss \@p@sfile \hss }
			\vss
			}}}
		}\else{
			\vbox to \@p@srheight true sp{
			\vss
			\hbox to \@p@srwidth true sp{\hss}
			\vss
			}
		}\fi

	}\fi
}}
\psfigRestoreAt

\begin{document}
\newcommand\sss{\scriptscriptstyle}
\newcommand\mq{\mbox{$m_{\sss \rm Q}$}}
\newcommand\mug{\mu_\gamma}
\newcommand\mue{\mu_e}
\newcommand\muf{\mu_{\sss F}}
\newcommand\mur{\mu_{\sss R}}
\newcommand\muo{\mu_0}
\renewcommand\pt{\mbox{$p_{\sss \rm T}$}}
\newcommand\as{\alpha_{\sss S}}
\newcommand\ep{\epsilon}
\newcommand\litwo{{\rm Li}_2}
\newcommand\aem{\alpha_{\rm em}}
\newcommand\refq[1]{$^{[#1]}$}
\newcommand\avr[1]{\left\langle #1 \right\rangle}
\newcommand\lambdamsb{
\Lambda_5^{\rm \sss \overline{MS}}
}
\newcommand\qqb{{q\overline{q}}}
\newcommand\asb{\as^{(b)}}
\newcommand\qb{\overline{q}}
\newcommand\sigqq{\sigma_{q\overline{q}}}
\newcommand\fqq{f_{q\qb}}
\newcommand\fqqs{f_{q\qb}^{(s)}}
\newcommand\fqqp{f_{q\qb}^{(c+)}}
\newcommand\fqqm{f_{q\qb}^{(c-)}}
\newcommand\fqqpm{f_{q\qb}^{(c\pm)}}
\newcommand\sigpq{\sigma_{\gamma q}}
\newcommand\mpq{{\cal M}_{\gamma q} }
\newcommand\fpq{f_{\gamma q}}
\newcommand\fpqs{f_{\gamma q}^{(s)}}
\newcommand\fpqp{f_{\gamma q}^{(c+)}}
\newcommand\fpqm{f_{\gamma q}^{(c-)}}
\newcommand\fpqpm{f_{\gamma q}^{(c\pm)}}
\newcommand\fpqtm{{\tilde f}_{\gamma q}^{(c-)}}
\newcommand\sigpg{\sigma_{\gamma g}}
\newcommand\mpg{{\cal M}_{\gamma g} }
\newcommand\fpg{f_{\gamma g}}
\newcommand\fpgs{f_{\gamma g}^{(s)}}
\newcommand\fpgm{f_{\gamma g}^{(c-)}}
\newcommand\fpgtm{{\tilde f}_{\gamma g}^{(c-)}}
\newcommand\siggg{\sigma_{gg}}
\newcommand\mgg{{\cal M}_{gg} }
\newcommand\fgg{f_{gg}}
\newcommand\fggs{f_{gg}^{(s)}}
\newcommand\fggp{f_{gg}^{(c+)}}
\newcommand\fggm{f_{gg}^{(c-)}}
\newcommand\fggpm{f_{gg}^{(c\pm)}}
\newcommand\fggtm{{\tilde f}_{gg}^{(c\pm)}}
\newcommand\fqg{f_{qg}}
\newcommand\epb{\overline{\epsilon}}
\newcommand\thu{\theta_1}
\newcommand\thd{\theta_2}
\newcommand\omxr{\left(\frac{1}{1-x}\right)_{\tilde\rho}}
\newcommand\omyo{\left(\frac{1}{1-y}\right)_{\omega}}
\newcommand\opyo{\left(\frac{1}{1+y}\right)_{\omega}}
\newcommand\lomxr{\left(\frac{\log(1-x)}{1-x}\right)_{\tilde\rho}}
\newcommand\MSB{$\overline{\mbox{MS}}$}
\newcommand\vltm{{\log\frac{-t}{m^2}}}
\newcommand\vlwm{{\log\frac{-u}{m^2}}}
\newcommand\vlpm{{\log\frac{1+\beta}{1-\beta}}}
\newcommand\vlpmq{{\log^2\frac{1+\beta}{1-\beta}}}
\newcommand\ltuno{{\log\frac{-t}{s}}}
\newcommand\ltdue{{\log\frac{-u}{s}}}
\newcommand\softb{{\litwo\frac{2\beta}{1+\beta}
                  -\litwo\frac{-2\beta}{1-\beta}
}}
%
\def\Cap{{\as\over{2\pi}}\,C_F}
\def\sw{\sin^2\theta_W}
\def\ee{e^+e^-}
\def\Ft{\tilde F}
\def\half{\mbox{\small $\frac{1}{2}$}}
\def\tird{\mbox{\small $\frac{1}{3}$}}
\def\ltap{\raisebox{-.4ex}{\rlap{$\sim$}} \raisebox{.4ex}{$<$}}
\def\gtap{\raisebox{-.4ex}{\rlap{$\sim$}} \raisebox{.4ex}{$>$}}
\def\VEV#1{\left\langle #1\right\rangle}
\def\LMS{\Lambda_{\overline{\mbox{\scriptsize MS}}}}
\def\rat#1#2{\mbox{\small $\frac{#1}{#2}$}}
\def\ltap{\raisebox{-.47ex}{\rlap{$\sim$}} \raisebox{.47ex}{$<$}}
\def\gtap{\raisebox{-.47ex}{\rlap{$\sim$}} \raisebox{.47ex}{$>$}}
\def\bom#1{\mbox{\bf{#1}}}
\def\beq#1{\begin{equation}\label{#1}}
\def\mbeq{\begin{equation}}
\def\beeq#1{\begin{eqnarray}\label{#1}}
\def\eeq{\end{equation}}
\def\eeeq{\end{eqnarray}}
\def \eq {e_{\sss Q}}
\def \ptg {\mbox{$p_{\sss T}^{Q\overline{Q}}$}}
\def \xf  {\mbox{$x_{\sss F}$}}
\def \dphi{\mbox{$\Delta\phi$}}
\def \dy  {\mbox{$\Delta y$}}
\def \pim {\mbox{$\pi^-$}}
\def \epem {\mbox{$e^+e^-$}}
\def \mc   {\mbox{$m_c$}}
\def \mb   {\mbox{$m_b$}}
\def \mqq   {\mbox{$M_{Q\overline{Q}}$}}
\def \tot   {{\rm tot}}
\def \nn    {\nonumber}
\renewcommand{\thefootnote}{\fnsymbol{footnote}}
\newcommand\qq{{\sss Q\overline{Q}}}
\newcommand\cm{{\sss CM}}
\input{psfig.sty}
\renewcommand\topfraction{1}       
\renewcommand\bottomfraction{1}    
\renewcommand\textfraction{0}      
\setcounter{topnumber}{5}          
\setcounter{bottomnumber}{5}       
\setcounter{totalnumber}{5}        
\setcounter{dbltopnumber}{2}       
%
\begin{titlepage}
\nopagebreak
\vspace*{-1in}
{\leftskip 11cm
\normalsize
\noindent
\newline
CERN-TH.7212/94 \newline
hep-ph/9404254 \newline
}
\vfill
\begin{center}
{\Large \bf\boldmath
Fragmentation Function Method for Charge Asymmetry Measurements in
$\ee$ Collisions}
\vfill
{\bf P.\ Nason\footnote{On leave of absence from INFN,
        Sezione di Milano, Milan, Italy.}}

        Theory Division, CERN, CH-1211 Geneva 23, Switzerland
\vskip .3cm
{\bf B.R.\ Webber\footnote{
Research supported in part by the U.K. Science and Engineering Research
Council and by the EC Programme ``Human Capital and Mobility", Network
``Physics at High Energy Colliders", contract CHRX-CT93-0537 (DG 12 COMA).}
}

        Cavendish Laboratory, University of Cambridge, UK.
\end{center}
\vfill
\nopagebreak
\begin{abstract}
{\small
We propose a method for measuring the hadron charge asymmetry in
$\ee$ collisions which is based upon the fragmentation function
formalism, and is largely independent of modelling of fragmentation
effects. Furthermore, in this method,
QCD radiative corrections can be accounted for
in a systematic way.
}
\end{abstract}
\vfill
CERN-TH.7212/94 \newline
April 1994    \hfill
\end{titlepage}
\section{Introduction}
Forward-backward
asymmetries measured at $\ee$ colliders near the $Z$
resonance can be used to measure the electroweak parameter
$\sin^2\theta_W$.
The differential cross section for the production of a light
fermion pair is given by
\beq{prima}
\frac{d\sigma_f}{d \cos\theta}=\sigma_f
\left(\frac{3}{8}(1+\cos^2\theta)+A^f_{\rm FB}(s)\cos\theta\right),
\eeq
where $\theta$ is the angle of the outgoing fermion $f$ with respect
to the direction of the incoming electron. The forward-backward
asymmetry is defined as
\beq{seconda}
A^f_{FB}=\frac{\sigma_f^F-\sigma^B_f}{\sigma_f^F+\sigma^B_f}
\eeq
where $\sigma^F_f$ ($\sigma^B_f$) is the cross section for producing
the fermion in the forward (backward) hemisphere. At leading order
on the $Z$ peak (neglecting photon channel contributions) it is
given by
\beq{terza}
A^f_{FB}(M^2_Z)=\frac{3}{4}\frac{2v_e a_e}{v^2_e+a^2_e}
                           \frac{2v_f a_f}{v^2_f+a^2_f}\;,
\eeq
where
\beq{quarta}
v_f=I^f_3-2e_f\sin^2\theta_W\, ,\quad\quad a_f=I^f_3\, ,
\eeq
with $e_f$ and $I_3$ denoting respectively the charge (measured
in units of the positron charge) and the third component of the
weak isospin.

Forward-backward asymmetries in the
production of leptons are measured at LEP and SLD, and they constitute
one very important measurement for electroweak tests
(see for example refs.~[\ref{SinthFromLeptons}]).
Due to the fact that $v_f$ is very
small for leptons, very high precision is required in this channel
in order to get a useful
measurement of $\sin^2\theta_W$. On the other hand, for
quarks the vector coupling is not small. There, however, one
has difficulty distinguishing between quarks of different
flavours, or even between quarks and antiquarks.
One approach is that of heavy flavour tagging in the hadronic final state
(see refs.~[\ref{SinthFromHeavy},\ref{HeavyTheory}]).
Another possibility is to use instead
the total charge asymmetry, averaged over all
quark flavours,
without flavour tagging. One defines
\beq{quinta}
A^{ch}_{FB}=\frac{1}{\Gamma_{had}}\left[
   \sum_{U=u,c}\Gamma_U A_{FB}^U-\sum_{D=d,s,b}\Gamma_D A_{FB}^D\right]
\eeq
and one then needs to find a final state quantity which is related to
the charge of the primary quark, at least on
a statistical basis.

Results from various approaches have been reported
in refs.~[\ref{ALEPH}-\ref{OPAL}].
The methods used there are all characterized by
some ``jet charge definition'', which should be used to assign a
charge to a jet. One then has to establish how often a jet with a
given charge comes from a quark with the same charge. This is computed
using Monte Carlo simulations. The drawback of this method is that it has
to rely upon the hadronization models built into the simulations.
A further drawback is that
it is not at all clear how the strong corrections to the produced
final state affect the measurement. One has no way to understand what
part of the radiative correction is already incorporated in the
simulation and what part is not. In fact, for a quantity
defined in this way, it may be 
impossible to define the radiative corrections properly.

In the present work, we propose a method for measurement of the charge
asymmetry which overcomes some of the above drawbacks: it does not
rely so heavily on hadronization models, and QCD radiative corrections
are fully and reliably calculable. Our method relies upon the 
reasonable assumption of valence dominance of the fragmentation
functions at large values of the energy fraction $x$, which can be
phrased in the following way:
at very large $x$, only the hadrons containing a primary quark survive.

In the following sections we give a detailed description of the
method, illustrated by some Monte Carlo results obtained using the
simulation program HERWIG\footnote{We used the current version, HERWIG 5.7,
with default parameter values.} [\ref{HERWIG}].

\section{Fragmentation function formalism}
The single inclusive cross section for the production of a charged hadron
is given by the formula
\beq{sigxth}
\frac{d^2\sigma^{h^\pm}}{dx\,d\cos\theta} =
{3\over 8} (1+\cos^2\theta) \frac{d\sigma_T^{h^+}}{dx} +
{3\over 4} \sin^2\theta \frac{d\sigma_L^{h^+}}{dx} \pm
{3\over 4} \cos\theta \frac{d\sigma_A^{h^+}}{dx} \; .
\eeq
We 
define the total and asymmetric
charged hadron fragmentation functions
\beqn
F^h&=&\int d\cos\theta \left( \frac{d^2\sigma^{h^+}}{dx\,d\cos\theta}
+\frac{d^2\sigma^{h^-}}{dx\,d\cos\theta} \right)=
2\left(\frac{d\sigma_T^{h^+}}{dx}+\frac{d\sigma_L^{h^+}}{dx}\right)
\nonumber \\
F^h_A&=&\int d\cos\theta \cos\theta
\left( \frac{d^2\sigma^{h^+}}{dx\,d\cos\theta}
-\frac{d^2\sigma^{h^-}}{dx\,d\cos\theta}
\right)=\frac{d\sigma^{h^+}_A}{dx}\;.
\eeqn
Following the notation of ref.~[\ref{NasonWebber}] we have
\beq{sesta}
F^h(x)=\sum_f\sigma_{0,f}(s)(D^{h^+}_f(x,s)+D^{h^-}_f(x,s))\;.
\eeq
We use the annihilation scheme, in which
the definition of the fragmentation function is such that the
coefficient functions for $F^h$ have no radiative corrections
if the scale in the fragmentation functions is set equal to
the centre-of-mass energy squared, $s$.
In the following we will drop the dependence on the scale,
assuming that it is always set equal to $s$.

In the annihilation scheme, the asymmetry coefficient function
does receive higher order corrections. We have
\beq{settima}
F^h_A(x)=\sum_f A_{f}(s)(\bar{D}^{h^+}_f(x)-\bar{D}^{h^-}_f(x))
\eeq
where
\beq{Dbarcorrected}
\bar{D}^h_f(x)=D^h_f(x)-\frac{C_F\as}{2\pi}\int_x^1 \frac{dz}{z}
(2-z)D^h_f(x/z)\;.
\eeq
This equation contains all that is needed to compute the
radiative corrections of order $\as$. For the moment, we will consider
only the leading order analysis, so we will assume that
$D^h_f=\bar{D}^h_f$.

The electroweak coefficients $\sigma_{0,f}(s)$ and $A_{f}(s)$
are given by the formulae
\beeq{fTA}
\sigma_{0,f}(s) &=& \frac{4\pi\alpha^2}{s}
[e_f^2 + 2e_f v_e v_f \rho_1(s) + (v_e^2+a_e^2)\,(v_f^2+a_f^2)\,\rho_2(s)]
\nonumber \\
A_f(s) &=& \frac{8\pi\alpha^2}{s}a_e a_f[e_f  \rho_1(s)
+ 2 v_e v_f \rho_2(s)]\; ,
\eeeq
where
\beeq{Rs}
\rho_1(s) &=& \frac{1}{4\sw\cos^2\theta_W}\cdot
\frac{s(m_Z^2-s)}{(m_Z^2-s)^2+m_Z^2\Gamma_Z^2}\nonumber \\
\rho_2(s) &=& \left(\frac{1}{4\sw\cos^2\theta_W}\right)^2
\frac{s^2}{(m_Z^2-s)^2+m_Z^2\Gamma_Z^2} \; .
\eeeq
On the $Z$ peak, neglecting the photon $s$-channel contribution, we have
\beqn\label{onpeak}
\sigma_{0,f}(s)&=&\frac{4\pi\alpha^2}{s}(v_e^2+a_e^2)(v_f^2+a_f^2)\rho_2(s)
\nonumber \\
A_f(s)&=&\frac{4\pi\alpha^2}{s}2 a_e v_e\,2 a_f v_f\; \rho_2(s)\;.
\eeqn
\section{Asymmetry from fragmentation functions}
Consider first the total and asymmetric fragmentation functions
for pions. According to our valence dominance hypothesis,
for large enough $x$ the $\pi^+$ can only come from a $u$ or a $\bar{d}$
quark.
We then have
\beqn
F^{\pi}(x)&=&\sigma_{0,u}(s)\,D^{\pi^+}_u(x)+
\sigma_{0,d}(s)\,D^{\pi^+}_{\bar{d}}(x) \nonumber \\
F_A^{\pi}(x)&=&A_{u}(s)\,D^{\pi^+}_u(x)-
A_{d}(s)\,D^{\pi^+}_{\bar{d}}(x)\;.
\eeqn
Using isospin symmetry and the on-peak expressions (\ref{onpeak}), we find
\beq{PionValDom}
\frac{ F_A^{\pi}(x) }{ F^{\pi}(x) }=
\left(\frac{2v_e a_e}{a_e^2+v_e^2}\right)\frac{ 2 a_u v_u - 2 a_d v_d}
{(a_u^2+v_u^2)+(a_d^2+v_d^2)}=-0.0375
\eeq
assuming $\sin^2\theta_W=0.23$, which we will use as our reference value.
The asymmetry is small in the pion case because of the partial
cancellation between the $u$ and $d$ contributions.

To test the validity of valence dominance in a particular hadronization
model, we studied the behaviour of the pion fragmentation functions using
HERWIG. In fig.~\ref{piratio} we plot $F_A^{\pi}(x)/F^{\pi}(x)$.
The error bars correspond to $2.5\times 10^5$
Monte Carlo events, i.e.\ about 5\% of the current LEP data.
The valence dominance value is also shown. We see that there
is agreement for $x>0.5$ at the 10 -- 20\% level. Corrections
of this order are not unexpected, for example due to
fast non-valence pions from resonance decays. It should be
possible to estimate such moderate corrections from hadronization
models without introducing a strong model dependence into the
determination of $\sin^2\theta_W$.  Recall that
$2v_e = 4\sin^2\theta_W - 1$, so that a 10\% precision in
eq.~(\ref{PionValDom}) implies about 1\% accuracy in $\sin^2\theta_W$.

\begin{figure}[htb]
\centerline{\psfig{figure=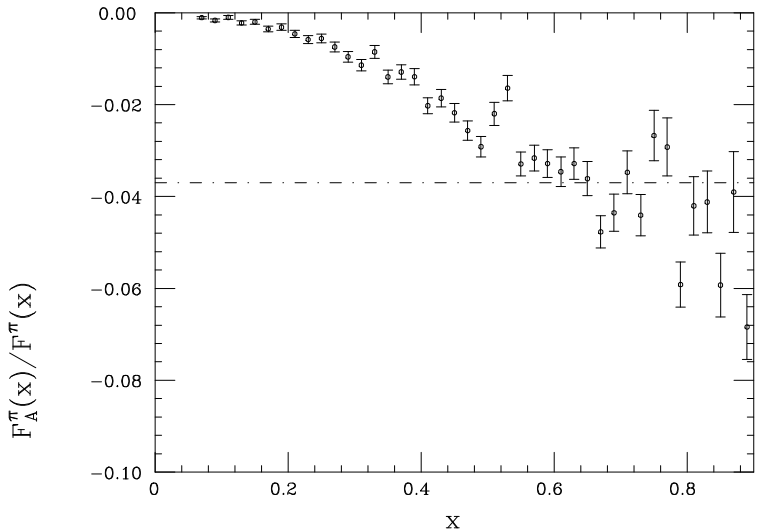,width=10cm,clip=}}
\caption[DATA]{ \label{piratio}
Ratio of asymmetry and total fragmentation functions for
charged pions, as a function of the energy fraction $x$.
The dot-dashed line shows the valence dominance
limit of eq.~(\ref{PionValDom}).
}
\end{figure}

Let us now consider the production of protons. Again using valence
dominance we get
\beqn
F^p(x)&=&\sigma_{0,u}(s)\left[D^{p}_u(x)+D^{p}_c(x)\right]+
\sigma_{0,d}(s)\left[D^{p}_{d}(x)+D^{p}_s(x)+D^{p}_b(x)\right] \nonumber \\
F_A^{p}(x)&=&A_{u}(s)\left[D^{p}_u(x)+D^{p}_c(x)\right]+
A_{d}(s)\left[D^{p}_d(x)+D^{p}_s(x)+D^{p}_b(x)\right]\;.
\eeqn
Observe that we have considered as valence dominant the production of
protons from strange, charm and bottom quarks. In fact, the production
of a strange, charm or bottom baryon will eventually lead through
weak decays to a stable nucleon, and in no case (in the valence
dominance regime) do we end up with an anti-baryon (baryon)
from an initial quark (anti-quark).
In the case of charm and bottom baryons,
these decays will lead to a considerable degradation of the initial
momentum. Since we intend to look at the large-$x$ region, let us
neglect for the moment the charm and bottom contributions.
We then get
\beq{ottava}
\frac{ F_A^{p}(x) }{ F^{p}(x) }=
\left(\frac{2v_e a_e}{a_e^2+v_e^2}\right)\frac{2 a_u v_u D^p_u(x)
+2 a_d v_d( D^p_d(x)+D^p_s(x) )}{
(a_u^2+v_u^2) D^p_u(x)+(a_d^2+v_d^2) ( D^p_d(x)+D^p_s(x) )}\;.
\eeq
Unlike the case of light mesons, baryon production through resonances
does not degrade the baryon momentum substantially. 
Resonance production and decay tend to randomize the charge of the
final nucleon. We may therefore, as a first approximation,
assume equality of the up, down and
strange fragmentation functions to produce protons at
large $x$, to obtain
\beq{PValDom}
\frac{ F_A^{p}(x) }{ F^{p}(x) }=0.1373\;.
\eeq
This value is much larger than in the pion case, since no cancellation
takes place. In fig.~\ref{protratio} we plot $F_A^{p}(x)/F^{p}(x)$.
As one can see, the Monte Carlo results approach the 
value (\ref{PValDom}) very quickly.
\begin{figure}[htb]
\centerline{\psfig{figure=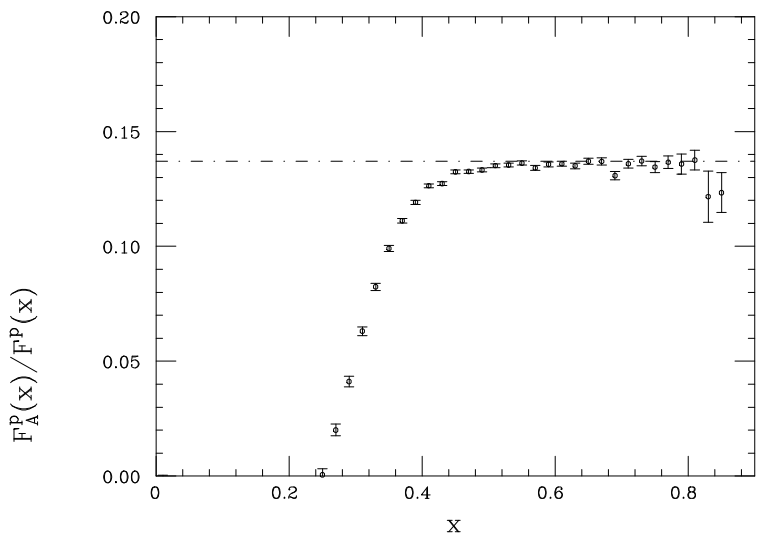,width=10cm,clip=}}
\caption[DATA]{ \label{protratio}
As in fig.~\ref{piratio}, but for protons and antiprotons.
Dot-dashed: eq.~(\ref{PValDom}).
}
\end{figure}

We can also obtain a larger asymmetry in the mesonic case by considering
pions and kaons together. We have in the valence
dominance approximation (again neglecting the mesons coming from
heavy flavour decays)
\beq{pikformula}
\frac{ F_A^{m}(x) }{ F^{m}(x) }=
\left(\frac{2v_e a_e}{a_e^2+v_e^2}\right)\frac{2 a_u v_u D^m_u(x)
-2 a_d v_d( D^m_{\bar{d}}(x)+D^m_{\bar{s}}(x) )}{
(a_u^2+v_u^2) D^m_u(x)+(a_d^2+v_d^2) ( D^m_{\bar{d}}(x)+D^m_{\bar{s}}(x) )}
\eeq
where $m$ stands for both charged pions and charged kaons.
In this case we would not expect 
equality of the three fragmentation
functions involved. In fact, in the valence dominance limit
the only charged meson into which a down quark can fragment
is the $\pi^-$. By isospin symmetry, $D^{\pi^-}_d$ should
equal $D^{\pi^+}_u$. An up quark, on
the other hand, can also go into a $K^+$. We therefore expect
$D^{m^+}_u>D^{m^-}_d$.
Our Monte Carlo studies suggest, however,
that for $x>0.5$ this difference is less than 20\%.

In the fragmentation
of strange quarks, one does not expect much leading pion production.
There are however pions coming from neutral kaon decays, which enhance
the charged strange fragmentation function without contributing to the
charge asymmetry, as assumed in formula~(\ref{pikformula}).
These effects merit further study,
both experimentally and theoretically.
For now we will make the rough assumption that the three fragmentation
functions are all equal to a first approximation.
Under this assumption we obtain
\beq{PiKValDom}
\frac{ F_A^{m}(x) }{ F^{m}(x) }\approx\left(\frac{2v_e
a_e}{a_e^2+v_e^2}\right)
\frac{2 a_u v_u-4 a_d v_d}
{(a_u^2+v_u^2)+2(a_d^2+v_d^2)}=-0.078\;,
\eeq
i.e.\ about twice the value in the case of pions alone.
Fig.~\ref{pikratio} shows a comparison with Monte Carlo results.

\begin{figure}[htb]
\centerline{\psfig{figure=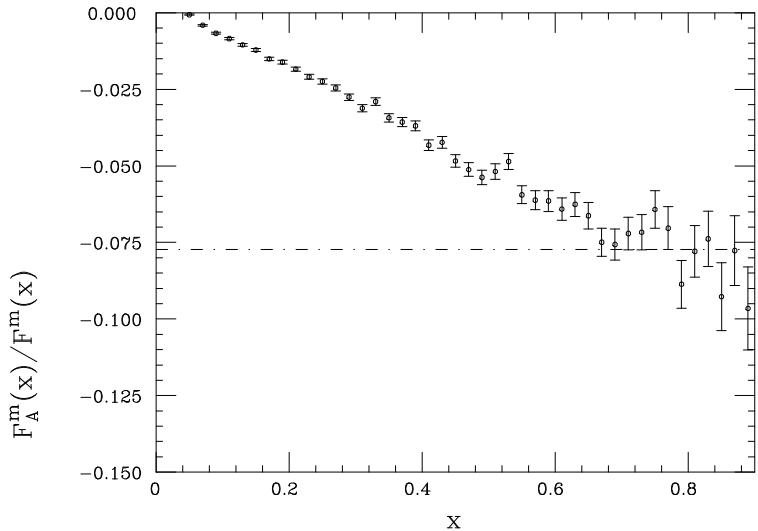,width=10cm,clip=}}
\caption[DATA]{ \label{pikratio}
As in fig.~\ref{piratio}, but for charged pions plus kaons.
Dot-dashed: eq.~(\ref{PiKValDom}).}
\end{figure}

For completeness, in Fig.~\ref{pikplus} we also show the
total and asymmetric charge meson fragmentation functions $F^{m}(x)$ and
$F_A^{m}(x)$. The values for $F_A^m$ are divided by the
limiting value of eq.~(\ref{PiKValDom}), to show the relation between
the two functions at large $x$. The dashed curve shows the result of
applying the higher order correction (\ref{Dbarcorrected}) to the
total fragmentation function.  The effect is small compared with
the probable magnitude of the corrections to valence dominance.

\begin{figure}[htb]
\centerline{\psfig{figure=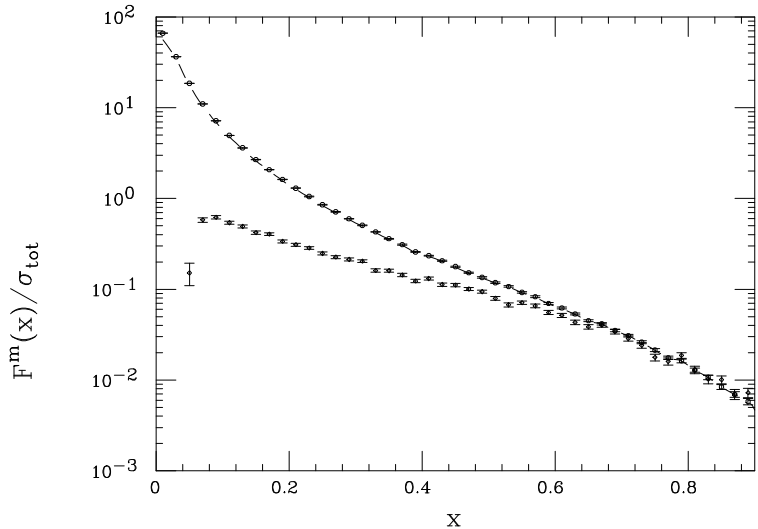,width=10cm,clip=}}
\caption[DATA]{ \label{pikplus}
Monte Carlo results on the asymmetry and total fragmentation functions
for charged mesons, (pions plus kaons), as
functions of the energy fraction $x$, at $E_{c.m.} = 91.2$ GeV.
The asymmetry has been divided by its
valence dominance limiting value of eq.~(\ref{PionValDom}).
Dashed curve: result of applying the higher order correction
(\ref{Dbarcorrected}) to the total fragmentation function.
}
\end{figure}
\section{Conclusions}
The method we propose for studying hadronic charge asymmetries, and
hence measuring $\sin^2\theta_W$, is based directly on the fragmentation
function formalism, for which the machinery of perturbative QCD is
already well developed. In particular the relation between the
total and asymmetric fragmentation functions for individual flavours
is known, and measurements at different energies can be related, to
next-to-leading order in $\as$. This makes the method attractive from
the theoretical viewpoint.

For phenomenological applications, one
needs to exploit relationships between the fragmentation functions
of quarks and their flavours.  The type of valence dominance
we have assumed seems less model dependent than other relationships
between quark and jet properties that have been proposed for this
purpose. Typical corrections are at the 10--20\% level at high $x$
and could be safely estimated from models. The predicted asymmetries
for protons and light mesons at the $Z$ resonance are substantial
and of opposite sign, providing a valuable cross-check on the
method.

Other cross-checks may be obtained experimentally.
One can, for example, require one fast particle in one jet,
and then study the fragmentation functions of the opposite jet.
For example, if one requires a fast $\pi^+$
on one side, (which, according to valence dominance, can only come
from a $u$ or a $\bar{d}$ quark),
this implies that on the opposite side we can measure
the linear combination of fragmentation functions
 $(v_u^2+a_u^2)D^h_{\bar{u}}+(v_d^2+a_d^2)D^h_d$ directly, for any $x$
value. This in turn can be used to test the valence dominance
hypothesis itself, and to measure the $K$ and $p$ fragmentation
functions for this combination. This measurement can in turn be used to
correct for deviations from the valence dominance hypothesis.
Many other combinations (for example requiring a fast proton or
$\Lambda_s$ or a kaon on one side) can be studied, allowing for a
wide variety of consistency checks and error estimates which are
independent of 
hadronization models.

We have concentrated here on light-quark hadrons, but of course the
method can also be applied to heavy quark fragmentation. In
particular the asymmetric and total $D^*$ fragmentation functions
at large $x$ would provide a rather clean measurement based on
charmed quarks.
\par \vskip 3mm
\noindent{\bf Acknowledgements}

We acknowledge useful discussions with
P.~Antilogus, A.~Blondel, D.~Rees, P.~Perrodo and R.~Jones.

\par \vskip 3mm
\noindent{\bf References}
\begin{enumerate}
\item \label{SinthFromLeptons}
ALEPH Collab., D.\ Buskulic et al., preprints CERN-PPE/93-40,
CERN-PPE/94-30;\newline
DELPHI Collab., P.\ Abreu et al., \zp{C59}{93}{21};
preprint CERN-PPE/94-31;\newline
L3 Collab., B.\ Adeva et al., \zp{C51}{91}{179};
preprint CERN-PPE/93-68;\newline
OPAL Collab., R.\ Akers et al., \zp{C61}{94}{19}.
\item \label{SinthFromHeavy}
L3 Collab., O.\ Adriani et al., \pl{B292}{92}{454};
preprint CERN-PPE/93-68; \\
OPAL Collab., P.D.\ Acton et al., \zp{C60}{93}{601}; \\
TOPAZ Collab., E.\ Nakano et al., \pl{B314}{93}{471}.
\item\label{HeavyTheory}
  A.\ Djouadi, J.H.\ K\"uhn and P.M.\ Zerwas,
  \zp{C46}{90}{441}; \\
  G.\ Altarelli and B.\ Lampe, \np{B391}{93}{3}; \\
  B.\ Lampe, preprint MPI-Ph/93-74.
\item \label{ALEPH}
ALEPH Collab., D. Decamp et al., \pl{B259}{91}{377}.
\item \label{DELPHI}
DELPHI Collab., P. Abreu et al., \pl{B277}{92}{371}.
\item \label{OPAL}
OPAL Collab., P.D. Acton et al., \pl{B294}{92}{436}.
\item \label{HERWIG}
  G.\ Marchesini and B.R.\ Webber, \np{B310}{88}{461};\\
  G.\ Abbiendi, I.G.\ Knowles, G.\ Marchesini, B.R.\ Webber,
  M.H.\ Seymour and \\
  L.\ Stanco, {\em Computer Phys.\ Comm.\ }
  {\bf 67}(1992)465.
\item \label{NasonWebber}
  P.\ Nason and B.R.\ Webber, preprint CERN-TH.7018/93, to be
  published in Nucl.\ Phys.\ B.
\end{enumerate}

\end{document}